\documentstyle[pra,aps]{revtex}

\begin{document}

\preprint{LYCEN 9507}
\draft
\title{Two-Photon Spectroscopy Between States of Opposite Parities}

\author{M. Daoud and M. Kibler}

\address{Institut de Physique Nucl\'eaire de Lyon, IN2P3-CNRS et
         Universit\'e Claude Bernard, F-69622 Villeurbanne Cedex,
         France}

\date{\today}

\maketitle

\begin{abstract}

Magnetic- and electric-dipole two-photon absorption (MED-TPA), recently
introduced as a new spectroscopic technique for studying transitions between
states of opposite parities, is investigated from a theoretical point of
view. A new approximation, referred to as {\it weak quasi-closure
approximation}, is used together with symmetry adaptation techniques
to calculate the transition
amplitude between states having well-defined symmetry properties. Selection
rules for MED-TPA are derived and compared to selection rules for
parity-forbidden electric-dipole two-photon absorption (ED-TPA).

\end{abstract}

\pacs{PACS: 42.65.$-$k, 71.35.$+$z
\hfill IPNL preprint LYCEN 9507}

\narrowtext

\centerline{\bf PRELIMINARIES}

\medskip

Two-photon spectroscopy is now an experimental technique widely used in
various domains, as for instance transition ions in
crystals and excitons in semiconductors and insulators. Two-photon
spectroscopy experiments
on excitons were first achieved by Hopfield and Worlock
\cite{hw}. Recently, Fr\"ohlich {\it et al}.~\cite{fipl} reported two-photon
absorption measurements for the three alkali halides RbI, NaI, and NaBr.
These authors considered nonlinear processes where two photons are
simultaneously absorbed,
one photon by magnetic-dipole transition and the other
           by electric-dipole transition. The resulting magnetic- and
electric-dipole two-photon absorption (MED-TPA) has to be distinguished from:

(i)
    The {\it classical} electric-dipole two-photon absorption (ED-TPA)
    where the two photons are simultaneously absorbed by electric-dipole
    transition between states of the same parity. The standard theory
    for parity-allowed ED-TPA was given by Axe \cite{axe} and the corresponding
    selection rules arising from the point symmetry group $G$ of the absorbing
    site were derived by Inoue and Toyazawa \cite{it} and by
    Bader and Gold \cite{bg}. Further investigations of parity-allowed ED-TPA
    were conducted in Refs.~\cite{st,kg,k9192,sdk92,dk93,kd93} on the basis of
    microscopic models and symmetry adaptation methods \cite{k79}
    for the chain of groups SU$(2) \supset G^*$
    (where $G^*$ is the double group of the group $G$).

(ii)
    The {\it forced} ED-TPA where the two photons are simultaneously absorbed
    by electric-dipole transition between states of opposite parities.
    Several parity violation mechanisms were introduced
    \cite{GHB,L87,MKY,SS} for explaining parity-forbidden ED-TPA, especially
    for partly-filled shell ions in crystals. Selection rules
    based on the  SU$(2) \supset G^*$  symmetry were
    obtained in Refs.~\cite{dk93,kd93,dk92} for parity-forbidden ED-TPA.

Of course, MED-TPA and (parity-allowed and parity-forbidden) ED-TPA
differ as far as simple considerations on spin are concerned. We may
also {\it a priori} expect some differences regarding the
selection rules coming from the point symmetry of the site.
In this connection, the selection rules for MED-TPA used in Ref.~\cite{fipl}
were obtained in the spirit of the
pioneer work by Inoue and Toyazawa \cite{it}. However, it is to be realized
that the selection rules introduced in Ref.~\cite{it} and extended by Bader and
Gold \cite{bg} were developed, on the basis of symmetry considerations only,
for (parity-allowed) ED-TPA.

It is the aim of this paper to show that the
SU$(2) \supset G^*$ selection rules for parity-forbidden ED-TPA hold
for MED-TPA when the two absorbed photons in both processes are different. We
shall show that when the two absorbed photons are identical (same energy, same
polarization, and same wave vector), there exists some differences,
besides the selection rules on spin, between the
selection rules for  MED-TPA  and  parity-forbidden  ED-TPA.
We shall illustrate these matters on
the $O_h$ symmetry considered in Ref.~\cite{fipl} for the paraexcitons in RbI,
NaI, and NaBr. The approach followed in the present work is not restricted to
qualitative arguments. It relies on the formalisms developed in Refs.~[7-11,17]
for intra- and inter-configurational two-photon transitions.
The precise selection rules for MED-TPA shall be derived from
a new approximation, which is less severe than the
closure approximation used by Axe \cite{axe} for ED-TPA, and from a
systematic use of symmetry adaptation techniques developed \cite{k79}
in the framework of the Wigner-Racah algebra for a chain
SU$(2) \supset G^*$. As a by-product, of great importance in two-photon
spectroscopy of transition ions in crystalline environments, the weak
quasi-closure approximation introduced in the present work is applied
to parity-allowed ED-TPA.

\bigskip

\centerline{\bf MAGNETIC- AND}
\centerline{\bf ELECTRIC-DIPOLE TRANSITIONS}

\medskip

We begin with MED-TPA (parity-allowed) transitions between states of opposite
parities. We first consider the case of two identical absorbed photons. From
the second-order time-dependent perturbation theory, the transition matrix
element between an initial state $i$ and a final state $f$ is
   \begin{eqnarray}
   M_{i \rightarrow f} & = &
   \sum_\ell
   {(f    | {\bf b} . {\bf M} | \ell )
    (\ell | {\bf e} . {\bf D} |i     ) \over
   E_i - E_\ell + \hbar \omega}                     \nonumber \\
                       & + &
   \sum_\ell
   {(f    | {\bf e} . {\bf D} | \ell )
    (\ell | {\bf b} . {\bf M} |i     ) \over
   E_i - E_\ell + \hbar \omega},                    \label{mif1}
   \end{eqnarray}
   which reflects the fact that when one photon is absorbed by magnetic-dipole
transition the other is absorbed by electric-dipole transition. In Eq.~(1),
${\bf e} . {\bf D}$ is the scalar product of the polarization vector
${\bf e}$ of the absorbed photons and of the electric dipolar operator
${\bf D}$ while
${\bf b} . {\bf M}$ is the scalar product of the vector
${\bf b} = {\bf k}_0 \times {\bf e}$
         (${\bf k}_0$ is the unit wave vector of
the radiative field) and of the magnetic dipolar operator
${\bf M}$. We take the initial and final state vectors in the form
  $|i) = |\alpha  J  a  \Gamma  \gamma )$ and
  $|f) = |\alpha' J' a' \Gamma' \gamma')$, where $\Gamma$ and $\Gamma'$ are
  irreducible representation classes (IRC's)
  of the group $G^*$. (We follow the notations of
  [7-12] for the symmetry labels $a \Gamma \gamma$ and  $a' \Gamma' \gamma'$
  of the states $i$ and $f$.)
  Similarly, the state vectors for the intermediate states $\ell$ are
  $| \ell ) = | \alpha '' J '' a '' \Gamma '' \gamma '')$. For the three
  kinds of
  state vectors, we neglect the $J$-mixing which may result, for example, from
  crystal-field effects. (The extension to the case where the $J$-mixing is
  taken into consideration is trivial: It is sufficient to proceed as in
  [7-11,17].)
The energy denominators in (1) have their usual meaning. In order to carry out
the summations over $\ell$ in (1), we adopt the approximation that the energies
$E_{\ell}$ are independent of the generalized magnetic quantum numbers
$a'' \Gamma'' \gamma''$. Therefore, let us put
    \begin{equation}
    E_{\ell} = E(\alpha'' J'').
    \label{wqca}
    \end{equation}
    This yields a weak quasi-closure approximation that is less restrictive
that the closure approximation used by Axe \cite{axe} for ED-TPA
[the latter approximation amounts to replace
$E_{\ell} = E(\alpha'' J'' a'' \Gamma'' \gamma'')$
by the barycenter of the intermediate states].
Then, by applying coupling and recoupling techniques for the chain
SU$(2) \supset G^*$ \cite{k79}, Eq.~(1) is amenable to the form
  \begin{eqnarray}
  M_{i(\Gamma  \gamma ) \rightarrow
     f(\Gamma' \gamma')}
                      & = & \sum_{k} \sum_{a'' \Gamma'' \gamma''}
  Y_k \big ( \alpha J, \alpha' J' \big )
  \big \{ {e} {b} \big \}^k_{a''\Gamma'' \gamma''} \nonumber \\
                      &   &
  \times
  f\pmatrix{
  J & J' & k\cr
  a\Gamma \gamma & a' \Gamma' \gamma' & a'' \Gamma'' \gamma''\cr
  }^*,
  \label{mif2}
  \end{eqnarray}
  where the parameters $Y_k$ are given by
\begin{eqnarray}
& &
 Y_k (\alpha J, \alpha' J') =
 (-1)^{2J'} \big [ k \big ]^{{1 \over 2}} \sum_{\alpha'' J''}
 \frac{1}{E_i - E(\alpha'' J'') + \hbar \omega}               \nonumber \\
& &
 \times
 \big [ (-1)^k  (\alpha'  J'  || M^1 || \alpha'' J'' )
                (\alpha'' J'' || D^1 || \alpha     J )        \nonumber \\
& &
             +  (\alpha'  J'  || D^1 || \alpha'' J'' )
                (\alpha'' J'' || M^1 || \alpha   J   ) \big ]
\left \{ \matrix{ J & k   & J'\cr
                  1 & J'' & 1 \cr } \right \}.                \nonumber \\
\label{yk}
\end{eqnarray}
Equation (4) should be understood as comprizing two sums over
$\alpha'' J''$: The first (respectively, second) sum is
to be expanded over states having the same parity
as the final (respectively, initial) state. The $f$ coefficients
in (3) are essentially Clebsch-Gordan coefficients
for SU$(2) \supset G^*$ since \cite{k79}
   \begin{eqnarray}
   & &
   f \pmatrix{
   j_1                   & j_2                   & k               \cr
   a_1 \Gamma_1 \gamma_1 & a_2 \Gamma_2 \gamma_2 & a \Gamma \gamma \cr }
   \nonumber \\
   & &
   =
   (- 1)^{2k} [j_1]^{-1/2}
   (j_2 k a_2 \Gamma_2 \gamma_2 a \Gamma \gamma | j_1 a_1 \Gamma_1 \gamma_1)^*.
   \label{f-Cgc}
   \end{eqnarray}
   The selection rules for MED-TPA easily follow from Eqs.~(3)-(5). The
  possible values of $k$ in (3) are $k=1$ and $2$. The value $k=0$ cannot
  contribute since ${\bf e}$ and
                   ${\bf b}$ are perpendicular
  (remember that
  $\{ {e} {b} \}^0  \sim  {\bf e} . {\bf b}$).
Furthermore, the possible values of $\Gamma''$ in (3) are given by the
existence
conditions of the $f$ coefficients. These conditions depend not only on the
point symmetry group $G$ but also on the rotation group O(3). To be more
precise, in the case where the two photons are identical, the selection rules
for the MED-TPA amplitude are the following:
(i)   $k = 1$ and $2$;
(ii)  the IRC $\Gamma''$ of $G$ is contained in the direct product
${\Gamma'}^* \otimes {\Gamma}$ as well as in the IRC ($k$) of O(3);
(iii) finally, parity selection rules
[involving parity symbols $u$ (or $-$) and $g$ (or $+$)]
have to be used together with (i) and (ii). The extension to the case
where the two photons are not identical leads to the results:
(i) $k=0$, $1$, and $2$; (ii) remains true; (iii) remains true.

The just
obtained selection rules for parity-allowed MED-TPA transitions may be
compared to the ones for parity-forbidden
ED-TPA transitions allowed between states
of different parities by a parity violation
mechanism. The transition moment $M_{i \rightarrow f}$
for such ED-TPA transitions is given [17] by a formula
similar to Eq.~(3) except that $\{ {e} {b} \}^k$ is replaced by
$\{ {e} {e} \}^k$ and the parameters $Y_k$ are generated by another
expression than Eq.~(4). Then, the selection rules for ED-TPA transitions
between states of opposite parities with identical (respectively, nonidentical)
photons are:
(i)   $k$ = 0 and 2 (respectively, $k$ = 0, 1, and 2);
(ii)  as for MED-TPA transitions;
(iii) as for MED-TPA transitions.
Therefore, the selection rules for ED- and MED-TPA transitions
turn out to be the same (except for spin) when
the two absorbed photons are not identical. However, a difference for the rule
(i) occurs when the two photons are identical. This provides us with a further
evidence that MED- and ED-TPA are two complementary spectroscopic techniques in
the case of identical photons.

\bigskip

\centerline{\bf AN ILLUSTRATIVE EXAMPLE}

\medskip

As an illustration, we consider the $G = O_h$ symmetry. For identical photons,
MED-TPA transitions $\Gamma   = \Gamma^+_1 \rightarrow
                     \Gamma ' = \Gamma^-_1$ are
forbidden (since the only possible
IRC $\Gamma '' = \Gamma^-_1$ requires that $k$ = 0),
in agreement with Ref.~[2]. On the other hand, MED-TPA transitions
$\Gamma  = \Gamma^+_1 \rightarrow
 \Gamma' = \Gamma^-_4$ are allowed
(since the only possible IRC $\Gamma '' = \Gamma^-_4$ requires $k = 1$),
in contradistinction to the assertion in Ref.~[2]. The
opposite situation occurs for parity-forbidden ED-TPA
transitions in the case of identical photons: The
$\Gamma^+_1 \rightarrow
 \Gamma^-_1$ transitions are allowed and the
$\Gamma^+_1 \rightarrow
 \Gamma^-_4$ transitions are forbidden.

Let us now illustrate, from a more quantitative point of view, the importance
of
the $k = 1$ contribution (which is not taken into account in Ref.~[2]) for
MED-TPA transitions with identical photons. By applying the formalism developed
in Refs.~[8-11], it is possible to get a closed-form expression for the
intensity
   \begin{equation}
   S(\Gamma \rightarrow \Gamma') =
   \sum_{\gamma \gamma '}
   \big | M_{i(\Gamma  \gamma ) \rightarrow
             f(\Gamma' \gamma')} \big |^2
   \label{inten}
   \end{equation}
   between the states of symmetry $\Gamma$ and $\Gamma'$
[the sum over $\gamma$ and $\gamma'$ in Eq.~(6) is extended
on all the Stark components of the initial and final states].
We have calculated
$S(\Gamma \rightarrow \Gamma')$ for the following situation: The wave vector
${\bf k}_0$ is taken along the crystallographic axis $(001)$
so that the polarization vector ${\bf e}$ has the components
${e}_0       = 0$ and
${e}_{\pm 1} = \mp {1 \over \sqrt 2} {\rm e}^{\pm {\rm i} \varphi}$
in linear polarization. Then Eq.~(6), used in
conjunction with the selection rules (i)-(iii) for MED-TPA transitions, can be
shown to yield
   \begin{eqnarray}
   S(\Gamma^+_1 \rightarrow \Gamma^-_3) & = & a \sin^2 2 \varphi, \nonumber \\
   S(\Gamma^+_1 \rightarrow \Gamma^-_5) & = & b \cos^2 2 \varphi, \nonumber \\
   S(\Gamma^+_1 \rightarrow \Gamma^-_4) & = & c,
   \label{3S}
   \end{eqnarray}
   where $a$, $b$, and $c$ are intensity parameters independent
of the polarization. The experimental situations considered
by Frvhlich {\it et al}.~[2] correspond to
${\bf e} =                   (100)$, i.e., $\varphi = 0            $, and
${\bf e} = {1 \over \sqrt 2} (110)$, i.e., $\varphi = {\pi \over 4}$. From
Eq.~(7), it is uncorrect to
assume that the transition observed in Ref.~[2]
for ${\bf e} = (100)$
is a transition to a
$\Gamma^-_5$ state because the transition to a
$\Gamma^-_4$ state has also a nonvanishing intensity.
For the same reason (viz., $c \not = 0$), it cannot
be assumed that the transition observed in Ref.~[2]
for ${\bf e} = {1 \over \sqrt{2}} (110)$
is a transition to a $\Gamma^-_3$ state. Consequently, a correct
assignment of the symmetry of the excitons considered in Ref.~[2]
requires some further polarization dependence experiments.
In this regard, it should be noted that in circular
polarization Eq.~(7) has to be substituted by
   \begin{equation}
   S(\Gamma^+_1 \rightarrow \Gamma^-_3)  =  a, \
   S(\Gamma^+_1 \rightarrow \Gamma^-_5)  =  b, \
   S(\Gamma^+_1 \rightarrow \Gamma^-_4)  =  0.
   \label{3SC}
   \end{equation}
   Then, the intensity of the $\Gamma^+_1 \rightarrow
                            \Gamma^-_4$ transition vanishes so that it
might be possible to get the anisotropic exchange splitting
$\epsilon_{\rm ex}$ (discussed in Ref.~[2])
between the $\Gamma^-_3$ and
            $\Gamma^-_5$ states
from measurements in circularly polarized light.

\bigskip

\centerline{\bf ELECTRIC-DIPOLE TRANSITIONS}

\medskip

An important facet that we would like to
briefly address in this paper concerns the
implication of the weak quasi-closure approximation, introduced above, on the
Axe [3] standard model for ED-TPA transitions between states of the same
parity.
Indeed, the use of the approximation (2) leads, for identical photons, to
the transition moment
  \begin{eqnarray}
  M_{i(\Gamma  \gamma ) \rightarrow
     f(\Gamma' \gamma')}
                      & = & \sum_{k} \sum_{a'' \Gamma'' \gamma''}
  Z_k \big ( \alpha J, \alpha' J' \big )
  \big \{ {e} {e} \big \}^k_{a''\Gamma'' \gamma''} \nonumber \\
                      &   &
  \times f
  \pmatrix{
  J               &                 J' &                     k \cr
  a \Gamma \gamma & a' \Gamma' \gamma' & a'' \Gamma'' \gamma'' \cr
  }^*,
  \label{mifEDED}
  \end{eqnarray}
  where
   \begin{eqnarray}
   & &
   Z_k (\alpha J, \alpha' J') =
   (-1)^{2 J'} [k]^{1/2}
   \sum_{\alpha '' J ''}
   \frac{1}{E_i - E(\alpha '' J '') + \hbar \omega}          \nonumber \\
   & &
   \times
   (\alpha  ' J ' || D^1 || \alpha '' J'')
   (\alpha '' J'' || D^1 || \alpha    J  )
   \left \{ \matrix{ J & k   & J'\cr
                     1 & J'' & 1 \cr } \right \}.            \nonumber \\
   \label{zk}
   \end{eqnarray}
   From Eqs.~(9) and (10), it is clear that the selection rules for
(parity-allowed) ED-TPA transitions with identical photons are:
(i)   $k = 0$ and 2;
(ii)  as      above;
(iii) as above. In the case of nonidentical photons, we have:
(i)   $k = 0$, 1, and 2;
(ii)  as          above;
(iii) as          above. The crucial difference
with the selection rule in Ref.~[3] is the occurrence of $k = 0$ (for both
identical or nonidentical photons). As a limiting case, when the energies
$E(\alpha'' J'')$ in Eq.~(10) are replaced by their barycenter, we recover the
quasi-closure approximation employed by Axe [3]. In this case, Eq.~(10)
gives back the results that only $k = 2$ contributes to Eq.~(9).

As an example, a parity-allowed ED-TPA transition
of type $\Gamma_1^{\pm} \rightarrow
         \Gamma_1^{\pm}$
(in any symmetry $G$) is not allowed in the Axe model
(since $\Gamma ' = \Gamma '' = \Gamma_1^{+}$ implies
that the only possible value of $k$ is $k = 0$).
On the contrary, such a transition becomes allowed if use is
made of the weak quasi-closure approximation inherent to Eq.~(2). It is to be
emphasized that the observation [18,19] of parity-allowed ED-TPA
transitions of type $J  = 0 \ (\Gamma_1^{\pm}) \rightarrow
                     J' = 0 \ (\Gamma_1^{\pm})$
was always attributed to the occurrence
of third- and higher-order mechanisms
  besides the second-order mechanisms
taken into account in the Axe model.
Our selection rules show that such transitions can be understood in the
framework of second-order mechanisms only once the severe approximation
$E_\ell = E(\alpha'' J'' a'' \Gamma'' \gamma'') = {\rm constant}$ is relaxed.
In this respect, the weak quasi-closure approximation based on Eq.~(2) is
phenomenologically equivalent to the introduction of third- and higher-order
mechanisms.

\bigskip

\centerline{\bf CONCLUSIONS}

\medskip

In conclusion, we have derived a model for MED-TPA
from the combination of (i) a weak quasi-closure
approximation for handling the G\"oppert-Mayer series
(1) and (ii) symmetry considerations based on the
group chain SU(2) $\supset G^*$. When the two absorbed photons
are identical, the obtained selection rules for MED-TPA
exhibit some important differences (besides the selection rule
on spin) with respect to those for ED-TPA between states of
opposite parities. The use of the weak quasi-closure
approximation in ED-TPA between states of the same parity leads to the
important result that a scalar contribution ($k=0$) may contribute to
the intensity in addition to the classical contributions
($k=2$ for identical photons and $k = 1,2$ for nonidentical photons).
It is to be observed that, for both ED- and MED-TPA, the structure
of the intermediate states is taken into account through the weak
quasi-closure approximation introduced in this paper. The selection rules for
MED-TPA have been discussed in connection
with the recent measurements on excitons by
Fr\"ohlich {\it et al}.~[2]; it is hoped that the  results in this work
[particularly Eqs.~(7) and (8)] would suggest some new MED-TPA experiments in
order to understand the discrepancy between the values for
$\epsilon_{\rm ex}$ for RbI obtained from MED-TPA and three-photon
absorption data.

\bigskip

\centerline{\bf ACKNOWLEDGMENTS}

\medskip

The authors are grateful to Prof.~D.~Fr\"ohlich for interesting
correspondence. Thanks are due to one of the Referees for a useful
comment.


\begin{references}

\bibitem{hw}
J.J. Hopfield and J.M. Worlock, Phys. Rev. {\bf 137}, A1455 (1965).

\bibitem{fipl}
D. Fr\"ohlich, M. Itoh, and Ch. Pahlke-Lerch, Phys. Rev. Lett. {\bf 72}, 1001
(1994).

\bibitem{axe}
J.D. Axe, Jr., Phys. Rev. {\bf 136}, A42 (1964).

\bibitem{it}
M. Inoue and Y. Toyazawa, J. Phys. Soc. Jpn. {\bf 20}, 363 (1965).

\bibitem{bg}
T.R. Bader and A. Gold, Phys. Rev. {\bf 171}, 997 (1968).

\bibitem{st}
G.E. Stedman, Adv. Phys. {\bf 34}, 513 (1985); see also:
G.E. Stedman, {\it Diagram Techniques in Group Theory}
(Cambridge Univ. Press, Cambridge, 1990).

\bibitem{kg}
M. Kibler and J.C. G\^acon, Croat. Chem. Acta {\bf 62}, 783 (1989).

\bibitem{k9192}
M.R. Kibler, in {\it Symmetry and Structural Properties of Condensed Matter},
edited by W. Florek, T. Lulek, and M. Mucha (World Scientific, Singapore,
1991);
see also: in {\it Excited States of Transition Elements}, edited by
W. Str\c ek, W.
Ryba-Romanowski, J. Legendziewicz, and B. Jez\.owska-Trzebiatowska (World
Scientific, Singapore, 1992).

\bibitem{sdk92}
J. Sztucki, M. Daoud, and M. Kibler, Phys. Rev. B {\bf 45}, 2023 (1992).

\bibitem{dk93}
M. Daoud and M. Kibler, J. Alloys and Compounds {\bf 193}, 219 (1993).

\bibitem{kd93}
M. Kibler and M. Daoud, Lett. Math. Phys. {\bf 28}, 269 (1993).

\bibitem{k79}
M.R. Kibler, in {\it Recent Advances in Group Theory and Their Application
to Spectroscopy}, edited by J.C. Donini (Plenum, New York, 1979).

\bibitem{GHB}
S.K. Gayen, D.S. Hamilton, and R.H. Bartram,
Phys. Rev. B {\bf 34}, 7517 (1986).

\bibitem{L87}
R.C. Leavitt,
Phys. Rev. B {\bf 35}, 9271 (1987).

\bibitem{MKY}
A.G. Makhanek, V.S. Korolkov, and L.A. Yuguryan,
Phys. Status Solidi (b) {\bf 149}, 231 (1988).

\bibitem{SS}
J. Sztucki and W. Str\c ek,
Chem. Phys. {\bf 143}, 347 (1990).

\bibitem{dk92}
M. Daoud and M. Kibler, Laser Phys. {\bf 2}, 704 (1992).

\bibitem{GJMBK}
J.C. G\^acon, B. Jacquier, J.F. Marcerou, M. Bouazaoui, and M. Kibler,
J. Lum. {\bf 45}, 162 (1990).

\bibitem{GBJKBA}
J.C. G\^acon, M. Bouazaoui, B. Jacquier, M. Kibler,
L.A. Boatner, and M.M. Abraham,
Eur. J. Solid State Inorg. Chem. {\bf 28}, 113 (1991).

\end{references}
\end{document}